\documentclass[aps,pra,reprint,showpacs,superscriptaddress,twocolumn]{revtex4-1}
\usepackage{amsmath}
\usepackage{amssymb}
\usepackage[dvipdf]{epsfig}
\usepackage{graphicx}
\usepackage{hyperref}
\usepackage[utf8x]{inputenc} 

\newcommand{\ket}[1]{\mbox{$|#1\rangle$}}

\begin{document}

\title{Superdiffusivity of quantum walks: 
  A Feynman sum-over-paths description} 

\author{F. M. Andrade}
\email{fmandrade@uepg.br}
\affiliation{
  Departamento de Matem\'{a}tica e Estat\'{i}stica,
  Universidade Estadual de Ponta Grossa,
  84030-900 Ponta Grossa-PR, Brazil
}

\author{M. G. E. da Luz}
\email{luz@fisica.ufpr.br}
\affiliation{
  Departamento de F\'{i}sica,
  Universidade Federal do Paran\'{a},
  C.P. 19044, 81531-980 Curitiba-PR, Brazil
}

\date{\today}

\begin{abstract}
Quantum walks constitute important tools in different applications, 
especially in quantum algorithms.
To a great extent their usefulness is due to unusual diffusive features, 
allowing much faster spreading than their classical counterparts.
Such behavior, although frequently credited to intrinsic quantum 
interference, usually is not completely characterized. 
Using a recently developed Green's function approach [Phys. Rev. A {\bf 84}, 
042343 (2011)], here it is described -- in a rather general way --
the problem dynamics in terms of a true sum  over paths history a la 
Feynman.
It allows one to explicit identify interference effects and also 
to explain the emergence of superdiffusivity.
The present analysis has the potential to help in designing quantum
walks with distinct transport properties.
\end{abstract}


\maketitle

\section{Introduction}

Quantum walks (QWs), a quantum version of classical random walks 
(CRWs) \cite{originally}, is a relatively simple class of systems, 
yet containing almost all the essential aspects of quantum mechanics 
\cite{kemp,wang}.
They can be used to model a large number of phenomena 
\cite{karski,broome}, such as:
energy transport in biological systems \cite{biology};
Bose-Einstein condensates redistribution \cite{chandrashekar1}; 
quantum phase transition in optical lattices \cite{chandrashekar2}; 
and decoherence processes \cite{ampadu}.
But, certainly where QWs have attracted more interest is in
quantum computing \cite{qw-qcomput,universal}.
In fact, QWs allow the development of new quantum algorithms 
\cite{ambainis}, which often display much better performance than 
their classical siblings \cite{ambainis,mosca}.

QWs usefulness in applications is in great part due to their unusual 
transport properties.
For instance, they present exponentially faster hitting times 
\cite{farhi,hitting-time,kempf} (the time necessary to visit any vertex 
in the system graph space), an important feature for searching in 
discrete databases \cite{searching1,searching2}.
Such faster spreading compared to CRWs \cite{kemp} is usually attributed 
to interference \cite{broome,oka-wojcik}, a key ingredient in 
implementations \cite{bowmeester-knight,rohde} and believed central to 
explain distinct behaviors \cite{kendon}.
However, exactly how these effects emerge in QWs usually is not totally 
characterized \cite{muelken}, posing challenges as to how one could
properly link the high degree of entanglement in QWs \cite{ide1} with 
interference.
Furthermore, since interference actually comes from a high proliferation 
of paths (after all, QWs are associated to the idea of CRWs \cite{qw-qrw}), 
it also bears on the problem of how decoherence \cite{kendon1} can
make the ``quantum trajectories'' to become classical \cite{brun1}.

In trying to understand interference in QWs, a path integral-like 
treatment would be appropriate.
Actually, a few interesting works along this line have been proposed 
\cite{yang-ide}.
However, they address the problem from a different perspective,
using combinatorial analysis to compute final states \cite{carteret-konno}, 
but not considering intermediary steps in terms of Feynman's history of 
trajectories \cite{feynman-book}.
Hence, interference is not made truly explicit.

In the present work we show how quantum interference determines QWs 
uncommon diffusive properties.
To this end, the exact Green's function \cite{andrade1} -- given as a general 
sum of paths -- is written in a closed analytical form.
Then, we describe how to calculate relevant quantities in a way 
identifying the trajectories superposition contributions.
To concretely illustrate the approach, we show that the usually observed: 
(a) complicated oscillatory behavior of the probability distribution for 
visits at different sites; and (b) the process dispersion dependence on 
time; are associated to the complex multiple reflections and transmission 
patterns of the system evolved paths.

Finally, we mention some important technical aspects.
There are several ways to formulate QWs, all defined in discrete
spaces (graphs) \cite{kemp}. 
Also, time may be either a continuous \cite{farhi} or a discrete variable.
In the latter, the major formulations are coined \cite{tregenna}
and scattering \cite{hillery} QWs.
Continuous time and coined are related through appropriate limits 
\cite{strauch-childs}, whereas coined and scattering are unitarily 
equivalent in any topology and for arbitrary transition 
amplitudes \cite{andrade2}.
Hence, we consider only scattering quantum walks (SQWs), keeping in mind
that our finds can be extended to such other constructions as well.
Moreover, the Green's function method considered here \cite{andrade1}
is valid for any graph topology. 
Although for our purposes we address QWs on the line, avoiding extra and 
unnecessary mathematical complications, we mention that the same type of 
analysis would likewise work in more complex networks.

\section{The sum over paths description}

\begin{figure}[t*]
\centerline{\psfig{figure=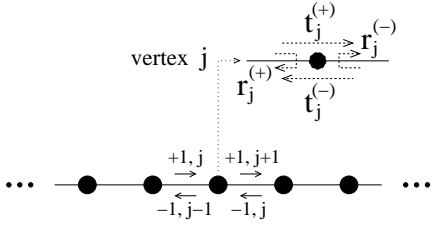,scale=0.29}}
\caption{QWs graph structure in 1D.
At each edge there are two basis states, e.g., $\ket{+, j}$ and 
$\ket{-,j-1}$ (schematically represented by arrows) for the edge 
between $j-1$ and $j$.
In detail the vertex dependent scattering quantum amplitudes.}
\label{fig:fig1}
\end{figure}

We assume an undirected 1D lattice of equally spaced vertices labeled 
in $\mathbb{Z}$, Fig. \ref{fig:fig1}.
Pairs of neighbor vertices are joined by a single edge.
To each edge we ascribe two basis states.
For instance, in Fig. \ref{fig:fig1} for the edge between $j-1$ and 
$j$ ($j$ and $j+1$) we have $\ket{+1, j}$ and $\ket{-1, j-1}$ 
($\ket{+1, j+1}$ and $\ket{-1, j}$).
Therefore, the full set $\{\ket{\sigma, j}\}$ spans all the possible system 
states $\ket{\psi}$. 
The quantum numbers $\sigma = \pm 1$ (hereafter for short $\pm$)
represent the propagation direction along the lattice (or graph).
The discrete dynamics is given by the one step time evolution operator 
$U$, such that the state at times $m+1$ and $m$ are related by
$\ket{\psi(m+1)} = U \ket{\psi(m)}$.
For an arbitrary phase $z = \exp[i \gamma]$, we have
\cite{andrade1} 
\begin{eqnarray}
  U^{\dagger} |\sigma, j \rangle
  &=& z^{*}
  \Big( 
  {t_{j-\sigma}^{(\sigma)}}^* \ket{\sigma,   j - \sigma } +
  {r_{j-\sigma}^{(-\sigma)}}^* \ket{-\sigma, j - \sigma } 
  \Big),
  \nonumber \\
  U |\sigma, j \rangle 
  &=& z 
  \Big( t_j^{(\sigma)} \ket{\sigma,  j + \sigma } +
  r_j^{(\sigma)} \ket{-\sigma, j - \sigma } \Big).
\end{eqnarray}
Here \cite{andrade2} (with $0 \leq r_j, t_j \leq 1$ and
$0 \leq \phi_{t,j}^{(\pm)},\phi_{r,j}^{(\pm)} < 2 \pi$)
\begin{eqnarray}
r_j^2 + t_j^2 = 1, &\qquad&
\phi_{r;j}^{(+)} + \phi_{r;j}^{(-)} = \phi_{t;j}^{(+)} + \phi_{t;j}^{(-)}
\pm \pi,
\nonumber \\
t_{j}^{(\pm)} = t_j \exp[i \phi_{t;j}^{(\pm)}], &\qquad&
r_{j}^{(\pm)}  = r_j \exp[i \phi_{r;j}^{(\pm)}],
\label{eq:condition-scattering}
\end{eqnarray}
guarantee the evolution unitarity.
The $r$'s and $t$'s can be understood as the vertices 
reflection and transmission quantum amplitudes (see Fig. \ref{fig:fig1}).

\begin{figure}[t*]
\centerline{\psfig{figure=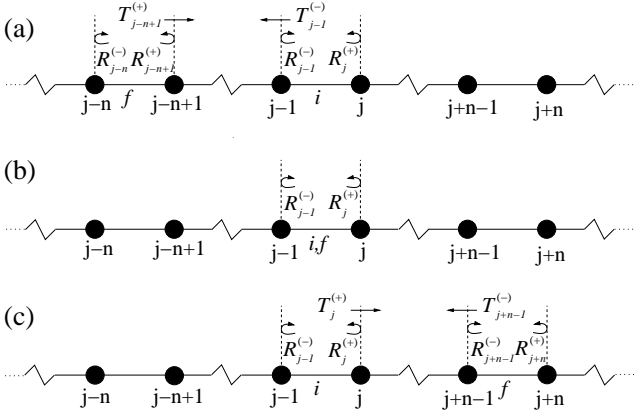,scale=0.35}}
\caption{For $\mathcal{G}$, the three possible 
situations for the relative positions of the initial and final, $i$
and $f$, edges.}   
\label{fig:fig2}
\end{figure}

The problem is fully described by the Green's function approach in 
\cite{andrade1}.
Consider the walk starting at the edge $i$ (between the vertices 
$j-1$ and $j$) with the state direction $\sigma$ and finally getting 
to the edge $f$ (between the vertices $j \pm n$ and 
$j \pm (n - 1)$, for any $n = 0, 1, 2, \ldots$).
The three possible situations are illustrated in Fig. 2.
Then, the most general exact expression for the Green's function,
representing the transmission probability amplitude 
$\{\sigma, i\} \rightarrow f$ reads ($s = -1 \ (+1)$
[in short $\mp$] for $f$ to the right (left) of $i$ and $s = 0$ for $f$ 
equal to $i$)

\begin{widetext}
\begin{equation}
\label{eq:green}
\mathcal{G}_{f, \{\sigma,i\}} =
\frac{ 
\left(z^{(3 + s \sigma)/2} \,
\left[
R_{j - (1-s)/2}^{(s)}\right]^{(1 + s \sigma)/2} T_{j-(s+1)/2}^{(-s)} 
\right)^{|s|} 
\left(1 + z \, R_{j - s n + (1-|s|)(\sigma-1)/2}^{(-s + (1-|s|) \sigma)} \right)}
{
\left(1 - z^2 \, R_{j - s n}^{(-s)} R_{j-s(n-1)}^{(s)} \right)^{|s|}
\left(1 - z^2 \, R_{j-1}^{(-)} R_{j}^{(+)} \right) - |s| \,
\left(z^4 \, R_{j+(s-1)/2}^{(s)} R_{j- s n}^{(-s)} T_{j-(s+1)/2}^{(-s)} T_{j-s(n-1)}^{(s)} 
\right)
}.
\end{equation}
\end{widetext}
The composed coefficients $R_{j}^{(\pm)}$ and $T_{j}^{(\pm)}$, functions
of the individual amplitudes $r_{j}^{(\pm)}$'s and $t_{j}^{(\pm)}$'s, are 
obtained from the following recurrence relations \cite{schmidt}
($\mu_{-} = j-(s+1)(n-1)/2$ and $\mu_{+} = j-1-(s-1)n/2$
for $s \neq 0$)
\begin{eqnarray}
R_{k}^{(\pm)} &=& r_{k}^{(\pm)} +
\frac{z^2 \, t_{k}^{(\pm)}  t_{k}^{(\mp)} R_{k\pm1}^{(\pm)}}
{1 - z^2 \, r_{k}^{(\mp)} R_{k\pm1}^{(\pm)}},
\ R_{\mu_\pm}^{(\pm)} = r_{\mu_\pm}^{(\pm)}, \nonumber \\
T_{k}^{(\pm)} &=&
\frac{z \, t_{k}^{(\pm)} T_{k\pm1}^{(\pm)}}
{1 - z^2 \, r_{k}^{(\mp)} R_{k\pm1}^{(\pm)}}, 
\ T_{\mu_\pm}^{(\pm)} = t_{\mu_\pm}^{(\pm)}.
\label{eq:recurrence}
\end{eqnarray}
In Eq. (\ref{eq:green}) it is not specified what is the 
final direction quantum number, $\nu$, when arriving at $f$.
In fact, it includes both cases once $\nu = \sigma \, (-\sigma)$ 
corresponds to the term $1$ ($z R$) in the second $(\ldots)$ in the 
numerator of Eq. (\ref{eq:green}).
There are different contexts for which we may seek the amplitude 
transition ${\{\sigma,i\} \rightarrow f}$.
Common ones are:
(i) exactly after $m = M$ time steps; and
(ii) when the system never visits vertices further to the left and 
to the right than, respectively, $j = J_{l}$ and $j = J_{r}$ (e.g.,
for first passage time calculations).
In both  we just need two extra relations for 
Eq. (\ref{eq:recurrence}): 
$R^{(+)}_{J_r} = r^{(+)}_{J_r}$ and $R^{(-)}_{J_l} = r^{(-)}_{J_l}$.
Moreover, for (i) we have $J_l = (j - 1) - 
[(M + s n - \delta_{1 \, s})/2]$ and 
$J_r = j + [(M - s n + \delta_{1 \, s})/2]$, with $[x]$ the integer 
part of $x$ and $n$ taken consistently.
Finally, for Eq. (\ref{eq:green}) obviously 
$\ket{\psi(0)} = \ket{\sigma, j + (\sigma-1)/2 }$.
For $\ket{\psi(0)} = \sum c_{\sigma,j} \ket{ \sigma, j}$, 
the correct Green's function would be
$\mathcal{G} = \sum c_{\sigma,j} \, 
\mathcal{G}_{f, \{\sigma, i|_{\sigma,j}\}}$.

The above exact expression is derived from a sum over infinite 
many ``scattering paths'' \cite{andrade1,luz}, starting and ending at 
the edges $i$ and $f$.
Its advantage is that all the possible quantum walk trajectories 
are ``compacted" into a closed formula.
So, distinct interference phenomena can be extract from 
$\mathcal{G}$.
Indeed, as demonstrated in \cite{andrade1}, this is achieved in a rather 
systematic way by means of two differential operators.
The probability for ${\{\sigma,i\} \rightarrow f}$
in exactly $m$ time steps is given by 
$p_{{\{\sigma,i\} \rightarrow f}}(m) = |\hat{S}_m \mathcal{G}_{f, \{\sigma,i\}}|^2$, 
with $\hat{S}_m = \frac{1}{m!} \frac{\partial^m}{\partial z^m}|_{z=0}$
the Step Operator.
To see it, we note that \cite{andrade1}
$\hat{S}_m \mathcal{G}_{f, \{\sigma,i\}} = \sum_{s.p.} \mathcal{P}_{s.p.}$,
for each $\mathcal{P}_{s.p.}$ being the contribution of a trajectory 
from $i$ to $f$ in $m$ steps \cite{note-0}.
Interference comes into play when we take the modulus square of such 
expansion.
Also, any specific ${\mathcal P}$ follows from 
$\hat{P}_{\mathcal P} \mathcal{G}$, for the Path Operator
(superscripts $\pm$ omitted for clarity)
\begin{equation}
\hat{P}_{\mathcal P} \, \cdot = 
\prod_{k,l \in \mathcal{P}}  
\frac{t_{k}^{m_{k}}}{m_{k}!} \,
\frac{r_{l}^{m_{l}}}{m_{l}!}
\bigg[\bigg(
\frac{\partial^{m_{k}}}{\partial t_{k}^{m_{k}}}
\frac{\partial^{m_{l}}}{\partial r_{l}^{m_{l}}}
\, \cdot
\bigg)\bigg|_{r_j,t_j=0, \forall j}
\bigg].
\label{eq:path-operator}
\end{equation}
In Eq. (\ref{eq:path-operator}) the $t_{k}$'s and $r_{l}$'s 
are the scattering amplitudes (appearing $m_{k}$ and $m_l$ times)
characterizing the path associated to $\mathcal{P}$ \cite{andrade1}.

With this mathematical `machinery', below we can make an analysis 
of sum over paths for QWs.

\section{Results and Discussion}

Suppose the initial state 
$\ket{\psi{(0)}}=\ket{\sigma,j}$, so we write 
\begin{equation}
U^{m}\ket{\psi(0)} = \ket{\psi(m)} = z^m
\sum_{j'=j-m}^{j'=j
+m} \ \sum_{\nu=\pm} \ a_{\nu,j'} \, \ket{\nu,j'}.
\label{eq:as}
\end{equation}
Above, some $a$'s are zero since certain basis states are absent, 
e.g., $|-, j+m-1\rangle$ cannot be reached in $m$ steps.
In fact, exactly $2 m$ $a$'s are not null.

Next, to make contact with CRWs, we observe that by leaving from the
edge corresponding to $\{\sigma,j\}$ there is a determined number of 
trajectories (eventually none) finally getting to specific edges in 
exact $m$ steps.
Thus, the total number of paths ending up in any possible $\ket{\nu,j'}$ is 
$2^m$.
Since $p = |a|^2$, the $a$'s can be given as the sum of the 
quantum amplitudes \cite{note-0} of all paths yielding 
$\{\sigma,j\} \rightarrow \{\nu, j'\}$.

\begin{figure}
\centerline{\psfig{figure=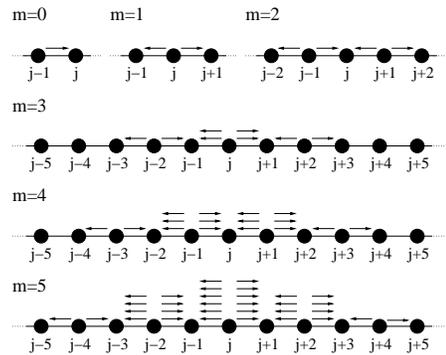,scale=0.2}}
\caption{For each $m$, an arrow $\rightarrow$ ($\leftarrow$) pointing 
to the vertex $j'$ indicates that the basis state $\ket{+,j'}$ 
($\ket{-,j'}$) is present in the expression for 
$|\psi(m)\rangle =  U^m \, \ket{+,j}$. 
The number of arrows of a given type equals the number of distinct paths 
leading to the corresponding $\ket{\nu,j'}$.}
\label{fig:fig3}
\end{figure}

As an simple example, for $|\psi(0)\rangle=\ket{+,j}$ the Fig. 
\ref{fig:fig3} schematically illustrates the basis states expansion
of $|\psi(m)\rangle$ (up to $m=5$).
Consider $m=2$, then
\begin{eqnarray}
z^{-2} \, |\psi(2)\rangle &=& z^{-2} \, U^{2} \ket{+,j} \nonumber \\
&=& a_{-,j-2} \, \ket{-,j-2} + a_{+,j} \, \ket{+,j} \nonumber \\
& & + 
    a_{-,j} \, \ket{-,j} + a_{+,j+2} \, \ket{+,j+2}. 
\label{example-fig3}
\end{eqnarray}
The presence of, say, $\ket{-,j-2}$ in the expression for $|\psi(2)\rangle$,
Eq. (\ref{example-fig3}), is represented in the $m=2$ case of Fig. 
\ref{fig:fig3} by an arrow pointing to the vertex $j-2$.
Moreover, its number, here just a single arrow, means there 
is only one path getting to $\ket{-,j-2}$ from $\ket{+,j}$ if $m=2$:
a trajectory which initially heading right at $j$ (since
$|\psi(0)\rangle=\ket{+,j}$), reverses its direction 
at $j$, goes to $j-1$ (first step), and finally goes to $j-2$ 
(second step), now heading left.
Note that in terms of a quantum scattering process, it represents a 
reflection from the vertex $j$ ($r_j^{(+)}$) and then a transmission through 
the vertex $j-1$ ($t_{j-1}^{(-)}$).

By applying $\hat{S}_m$ on $\mathcal{G}_{\{\nu, j'\},\{\sigma, j\}}$ 
and afterwards simply setting $r_j^{(\pm)} = t_j^{(\pm)} = 1$ \ $\forall j$,
one directly finds that the number of paths leading to
$\{\sigma,j\} \rightarrow \{\nu, j'\}$ after $m$ steps is
given by the binomial coefficient \cite{note}
\begin{equation}
  \mathcal{N}_{\nu,j'} =
  \binom{m-1}{\frac{m+j'-j}{2}-\delta_{\sigma \nu}}.
  \label{eq:num-paths}
\end{equation}
As a simple check, one can test Eq. (\ref{eq:num-paths}) with the 
schematics in Fig \ref{fig:fig3}.
Moreover, from the mapping between SQWs and coined QWs in \cite{andrade2}, 
the number of paths to a given $j'$ state for the latter QW formulation 
is trivially derived from $\mathcal{N}_{\nu,j'}$ as
\begin{equation}
\mathcal{N}_{j'}^{{\mbox{\scriptsize coined}}} = 
\sum_{\nu} \mathcal{N}_{\nu,j'} =
\binom{m}{\frac{m+j'-j}{2}} = \binom{m}{\frac{m-j'+j}{2}} ,
\label{eq:number_paths_cl}
\end{equation}
which agrees with the formula in Ref. \cite{brun2} (with
$j=0$). 

Assume any path taking, regardless the order, $d^{(-)}$ ($d^{(+)}$) 
steps to the left (right).
It would lead the system to $|\nu,j' = j + d^{(+)} - d^{(-)}\rangle$. 
Reversing this reasoning, consider a fixed
$m  = d^{(+)} + d^{(-)} \geq |\Delta j|$, with $\Delta j = j' - j$.
Paths for which $d^{(\pm)} = (m \pm \Delta j)/2$ are both integers will 
result in $j \rightarrow j'$.
To obtain all such paths, we should consider $\mathcal{G}$ for
$J_{l} = j - d^{(-)}$ and $J_{r} = j + d^{(+)}$.
The contribution from each path to a given coefficient in Eq. 
(\ref{eq:as}) will involve exactly $m$ position dependent amplitudes 
$r_j^{(\pm)}$'s and $t_j^{(\pm)}$'s.
In this way, the actual procedure to calculate the $a$'s is to
compute $a_{\nu, j'} = \hat{S}_m \mathcal{G}_{\{\nu, j'\},\{\sigma, j\}}$,
for $J_l$, $J_r$, $d^{(\pm)}$ as above.

For complete arbitrary $r_j^{(\pm)}$'s and $t_j^{(\pm)}$'s and for 
$m$ in the hundreds, any available computer algebra system can
be used to obtain the $a$'s as explained.
Actually, vertex-dependent quantum amplitudes can give rise to a great 
diversity of diffusive properties \cite{diffusion-position}.
So, the present procedure may be useful to test distinct QWs models 
transport features, helping to choose sets of $r_j^{(\pm)}$'s 
and $t_j^{(\pm)}$'s more appropriate in different applications (examples 
to appear elsewhere).

However, a real surprise is for the situation when superdiffusion 
takes place even for $j$ independent quantum amplitudes and when at 
each single step the QW resembles an unbiased classical walk (i.e.,
$50$--$50\%$ probability to go right-left) \cite{kemp}.
In the following, we show how interference can fully explain this 
apparently non-intuitive behavior.

For $r_j^{(\pm)} = r^{(\pm)}$, $t_j^{(\pm)} = t^{(\pm)}$ and from 
the above prescription, we get 
($n_{sup} = \min\{d^{(\sigma)}-\delta_{\sigma \nu}, d^{(-\sigma)}-1\}$)
\begin{eqnarray}
\label{eq:amplitudes}
a_{\nu,j'} &=& 
\sum_{n=-\delta_{\sigma \nu}}^{n=n_{sup}} f_n \, C_n, \ \ \ 
f_n = \binom{d^{(\sigma)}}{n + \delta_{\sigma \nu}}
\binom{d^{(-\sigma)}-1}{n}, \nonumber \\
C_n &=& [t^{(\sigma)}]^{d^{(\sigma)} - n - \delta_{\sigma \nu}} \;
[r^{(-\sigma)}]^{n + \delta_{\sigma \nu}} \nonumber \\
& & \times
[t^{(-\sigma)}]^{d^{(-\sigma)}-n-1} [r^{(\sigma)}]^{n+1}.
\end{eqnarray}
Furthermore, using Eq. \eqref{eq:condition-scattering}
\begin{equation}
\label{eq:amplitudes-rt}
C_n = \exp[i \phi] \; t^{m} \;
\left(\frac{r}{t}\right)^{2 n + \delta_{\sigma \nu} + 1}
\, (-1)^n,
\end{equation}
with $\phi$ a global phase (unimportant here) which depends on 
$j$, $j'$, $\sigma$, $\nu$, $m$ and $\phi_{r,t}^{(\pm)}$.
In Eq. \eqref{eq:amplitudes}, $f_n$ gives the number of distinct paths 
yielding a same amplitude $C_n$ to the $a$'s.
This is possible because different paths correspond to a different
order of scattering processes along the lattice.
Nevertheless, if the final set of scattering's coincides, 
the resulting amplitudes $C_n$ are equal.
The total number of paths for $a_{\nu, j'}$ is 
${\mathcal N}_{\nu,j'} = \sum_n f_n = \binom{m-1}{d^{(\sigma)} - 
\delta_{\sigma \nu}}$, which agrees with Eq. (\ref{eq:num-paths}).

Particularly important in Eq. (\ref{eq:amplitudes-rt}) is the 
factor $(-1)^{n}$, arising from the phases difference, Eq. 
(\ref{eq:condition-scattering}), 
between reflections and transmissions in a trajectory.
In fact, for each path the number of directions change along the way is
$2 n + 1 + \delta_{\sigma \nu}$.
Therefore, distinct paths may contribute with distinct signals (through 
$(-1)^{n}$) to the sum in Eq. (\ref{eq:amplitudes}), leading to 
constructive or destructive interference.

\begin{figure}[t]
\centering
\includegraphics*[width=0.48\textwidth]{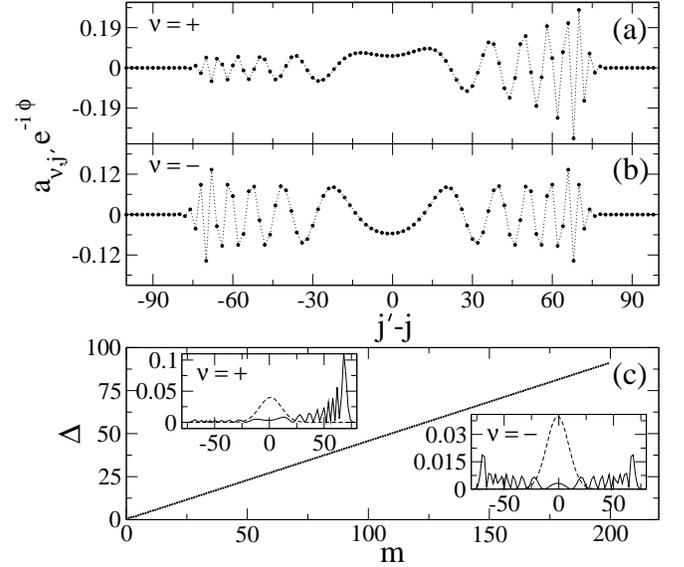}
\caption{Up to a global phase, the dimensionless coefficient 
$a_{\nu,j'}$, Eq. (\ref{eq:a-final}), as function of the quantum number
$j'$ for $|\psi(0)\rangle = |+,j\rangle$, $m=100$, (a) $\nu=+$, and (b) 
$\nu=-$. 
Since $m$ is even, $a_{\nu,j'} = 0$ if $j'- j$ is odd.
The notorious \cite{kemp} amplitudes asymmetry because the particular 
initial state arises only for $\delta_{\sigma \nu} = 1$ (case (a)).
(c) The linear dependence of the standard deviation $\Delta$ on the 
discrete time $m$.
In the insets $|a_{\nu,j'}|^2$ and the corresponding 
probabilities for an unbiased classical random walk (dashed curves)
vs. $j'- j$.}
\label{fig:fig4}
\end{figure}

Lastly, in the ``unbiased'' case of $r=t=1/\sqrt{2}$, i.e., $50\%$--$50\%$
reflection-transmission probability in each vertex (for a similar coined 
case see, e.g., \cite{brun2}), Eq. (\ref{eq:amplitudes}) reduces to
\begin{eqnarray}
\label{eq:a-final}
& & a_{\nu, j'} = \exp[i \phi] \, 2^{-m/2} \{
-2^{m} \delta_{m  \, d^{(\sigma)}} + 
[d^{(\sigma)}]^{\delta_{\sigma \nu}} \nonumber \\
& & \times _2F_1(-d^{(\sigma)} + \delta_{\sigma \nu}, 
-d^{(-\sigma)} + 1; 1 + \delta_{\sigma \nu}; -1)\},
\label{eq:hyper}
\end{eqnarray} 
with $_2F_1$ the Gaussian hypergeometric function.
To illustrate this formula, we consider the $a$'s for the final states 
$\ket{+,j+1}$ and $\ket{+,j+3}$ in the case $m=5$ of Fig. 
\ref{fig:fig3} (thus $\delta_{\sigma \nu} = 1$).
From Eq. (\ref{eq:hyper}) we get $a_{+,j+1} = 0$ and 
$|a_{+,j+3}|^2 = 1/2$.
To understand why, note that from $f_n$ in Eq. (\ref{eq:amplitudes}) or 
by inspecting Fig. \ref{fig:fig3}, we find there are six (four) possible 
paths leading to $|+,j+1\rangle$ ($|+,j+3\rangle$).
For $j'= j+1$, three paths have two ($n=0$) direction changes
and three have four ($n=1$).
The phases are then, respectively, $(-1)^{0}=1$ and $(-1)^{1}=-1$.
Hence, these two groups of paths suffer destructive interference.
On the other hand, for $j'=j+3$ there are four possible paths, 
all with two direction changes ($n=0$) and thus with a same phase.
The paths therefore build up a relatively high amplitude. 

The above results can also explain two typical and important 
behaviors observed in QWs \cite{kemp,kendon} (see Fig. \ref{fig:fig4} 
(a)-(c)):
(i) for usual CRWs, the probabilities for the particle location are 
Gaussian distributed, with a standard deviation of $\sqrt{m}$ (insets 
of Fig. \ref{fig:fig4} (c)).
On the other hand, quantum mechanically the $|a|^2$'s, representing the
particle distribution along the graph, are not spatially concentrated; 
(ii) $a_{j'}$ vs. $j'$ presents stronger oscillations for the
$j'$'s far away from the initial $j$, a pattern usually without a 
classical analog.

In fact, both (i) and (ii) originate from a similar mechanism.
For a fixed large $m$, the number of trajectories $\mathcal{N}_{\nu,j'}$ 
leading to $j'$ is large (small) if $|\Delta j|$ is small (large). 
In the classical case, since there are no interference, 
the probabilities are directly proportional to $\mathcal{N}_{\nu,j'}$
and the Gaussian distribution naturally emerges (recall that
binomial distributions, c.f. Eq. (\ref{eq:num-paths}), converge to 
Gaussians).
In the quantum case, the many cancellations coming from opposite signals 
for distinct groups of trajectories, Eq. (\ref{eq:amplitudes-rt}), 
prevents the probabilities at $|\Delta j|$ small to be much higher than 
those at larger $|\Delta j|$, Fig. \ref{fig:fig4} (a)-(b).
Hence, a more balanced distribution among the states $j'$'s is obtained.
By the same token, the smooth (strong oscillatory) behavior for 
$|\Delta j|$ small (large) is due to the fact that varying $j'$ in 
such interval will proportionally cause a small (large) change in 
the number of paths contributing to $a_{j'}$.
This results in a slow (rapid) variation of $a_j'$ as function of $j'$.

Thus, the observed system fast spreading, e.g., in the unbiased case
characterized by a linear dependence on $m$ for the standard 
deviation ($p_{j'} = |a_{+,j'}|^2 +  |a_{-,j'}|^2 $):
$\Delta = 
\sqrt{\sum_{j'} (j'-j)^2 p_{j'} - (\sum_{j'} (j'-j) p_{j'})^2}$
(Fig. \ref{fig:fig4} (c)); is due to (i)-(ii).
In their turn, (i)-(ii) are a direct consequence of intricate 
interference effects among paths with different phases.

\section{Conclusion}

Summarizing, our contribution here has been twofold.
First, we propose a distinct approach -- based on a true sum over paths 
history -- to study QWs in general.
It leads to some exact analytical results, which may be difficult to 
obtain by other means.
Second, we properly quantify a fundamental characteristic of QWs, 
interference, explicit associating such phenomenon with the emergence 
of supperdiffusive behavior.

Hence, the present framework provides a powerful tool to test 
distinct aspects of QWs evolution, and whose complete comprehension is 
certainly an important step towards making QWs more reliable to distinct 
applications as in quantum computing.

\section*{Acknowledgments}

MGEL acknowledges a research grant from CNPq.
Financial support is also provided by Finep/CT-Infra.

\end{document}